# Primary Initiation of Submarine Canyons

**J. Marvin Herndon**
**Transdyne Corporation**
**San Diego, CA 92131 USA**

mherndon@san.rr.com

**Abstract:** The discovery of close-to-star gas-giant exo-planets lends support to the idea of Earth's origin as a Jupiter-like gas-giant and to the consequences of its compression, including whole-Earth decompression dynamics that gives rise, without requiring mantle convection, to the myriad measurements and observations whose descriptions are attributed to plate tectonics. I propose here another, unanticipated consequence of whole-Earth decompression dynamics: namely, a specific, dominant, non-erosion, underlying initiation-mechanism precursor for submarine canyons that follows as a direct consequence of Earth's early origin as a Jupiter-like gas-giant.

The origin of submarine canyons, ubiquitous, prominent features of continental margins that sometimes in scale rival the Grand Canyon, has been the subject of much debate. Virtually all submarine canyons are subject to erosion, so the origin-debate generally focuses on the nature and circumstances of their formation by erosion-processes [1-3]. While there may be more than one submarine-canyon initiation-process, the commonality of their occurrence in diverse environments suggests the potential dominance of one underlying mechanism. The purpose of this brief communication is to propose that specific, dominant, underlying initiation-mechanism: an unanticipated, non-erosion precursor that follows as a direct consequence of Earth's early origin as a Jupiter-like gas-giant.

The presently popular idea of Solar System formation began with publication in 1963 of an assumption-based model by Cameron [4]. The idea, later called the "standard model of solar system formation", was that dust would condense from a diffuse, hot gas of solar composition at a pressure of about $10^{-5}$ bar. The dust would then collect into balls, then into rocks, then into boulder-size accumulations, and finally into planetesimals; these would eventually gather to form a planet like Earth. Subsequently, I showed from thermodynamic considerations why the model was wrong: Essentially all of the condensable elements of a gas of solar composition at such a low pressure would condense as oxides leaving no iron metal for a massive-core planet like Earth [5].

In 1944, Eucken [6] described from thermodynamic considerations Earth's formation by condensation at high pressures, ≥ 100 bar, in the interior of a hot gaseous protoplanet of solar composition. In such an environment, molten iron (and the elements dissolved therein), being the most refractory condensate, rain-out to form Earth's core before condensation forms the mantle. I have verified Eucken's calculations and extended his concept to include complete condensation of volatile elements [5, 7, 8]. Fully condensed with about 300 Earth-masses of hydrogen, helium, and other volatiles, Earth's gas-giant, pre-Hadean mass was virtually identical to that of Jupiter [5]. My idea of Earth having an early origin as a Jupiter-like gas giant is not at all strange as close-to-star gas-giant exo-planets are observed in other planetary systems [9].

The dynamics of planet Earth, a direct consequence its early gas-giant origin, is described by my new geodynamic theory, called *whole-Earth decompression dynamics* [10-12], which gives rise to the myriad measurements and observations whose descriptions are attributed to plate tectonics, but without requiring mantle convection. It provides a basis for Earth's previously having had reduced radius and a mechanism and powerful energy source for its decompression. Whole-Earth decompression dynamics obviates the primary pitfall of plate tectonics theory, the necessity of mantle convection, and overcomes the problems of endemic to Earth expansion theory, the absence of basis, energy source, and appropriate mechanism.

Envision pre-Hadean Earth, compressed to about 64% of present radius by about 300 Earth-masses of primordial gases and ices. At some point, after being stripped of its massive volatile envelope, presumably by the Sun's super-intense T-Tauri solar winds, internal pressures would build, eventually cracking the rigid crust. Powered by the stored energy of protoplanetary compression, Earth's progressive decompression is manifest at the surface by the formation of cracks: *primary* decompression cracks with underlying heat sources capable of extruding basalt, and *secondary* decompression cracks without heat sources that serve as ultimate repositories for basalt extruded from primary decompression cracks. Mid-ocean ridges and submarine trenches, respectively, are examples of these. Secondary decompression cracks serve to increase surface area in response to decompression-driven volume expansion. Basalt extruded at mid-ocean ridges becomes seafloor, spreading and eventually subducting, i.e., falling into secondary decompression cracks, seismically imaged as "down-plunging slabs", but without engaging in the process of mantle convection.

Whole-Earth decompression dynamics not only gives rise to the multitude of observations attributed to plate tectonics (and without mantle convection), but it affords understanding that transcends, being responsible for Earth's well-documented features: (1) Orogeny occurs not only by "plate collision", but by the vertical uplift intrinsic to decompression; (2) Partially in-filled secondary decompression cracks uniquely explain oceanic troughs, which are generally inexplicable by plate tectonics; and, (3) Compression heating at the base of the rigid crust is a direct consequence of mantle decompression, but has no parallel in plate tectonics. Here I present another extension by proposing dominant submarine-canyon-initiation as a consequence of whole-Earth decompression dynamics.

As illustrated in Figure 1, consider a hypothetical 4000 km "ancient" continent cross-section (arc *ABC*) at a radius of 64% of Earth's present radius, *OB,* and the same "present" continent cross-section (arc *DEF*) at present Earth radius, *OE*. Consider the line *OR* as a fixed reference, indicating that no "continental drift" has occurred, as the reference line *OR* bisects both the ancient and present continent. Observe that the ancient continent subtends an angle, <*AOC*, that is considerable greater than the angle the present continent subtends, <*DOF*: 56.3 degrees versus 36.0 degrees. Note that, although the length of the both the ancient and present surface continent cross-section are the same, 4,000 km, the chord lengths, indicated by dotted lines, differ: 3841.3 km versus 3934.5 km, respectively. Whole-Earth decompression causes changes in surface curvature that, I submit, are responsible for the primary initiation of submarine canyons.

Suppose that the coastline of the hypothetical continent, shown in Figure 1, is circular. Then the perimeter of its circular coast line is simply π times its chord length. Whole-Earth decompression, in the example above, necessitates an increase in peri-continental circumference of about 2.4%. The percent circumference-increase for other hypothetical continental arc lengths is shown in Figure 2.

Whereas rocks can undergo considerable compression, their brittle nature and low tensile strength precludes adjusting to the requisite increase in circumference caused by whole-Earth decompression without tension fractures occurring. Such decompression-generated, peri-continental tension-fractures, I posit, initiate the formation of submarine canyons; erosion-processes do the rest.

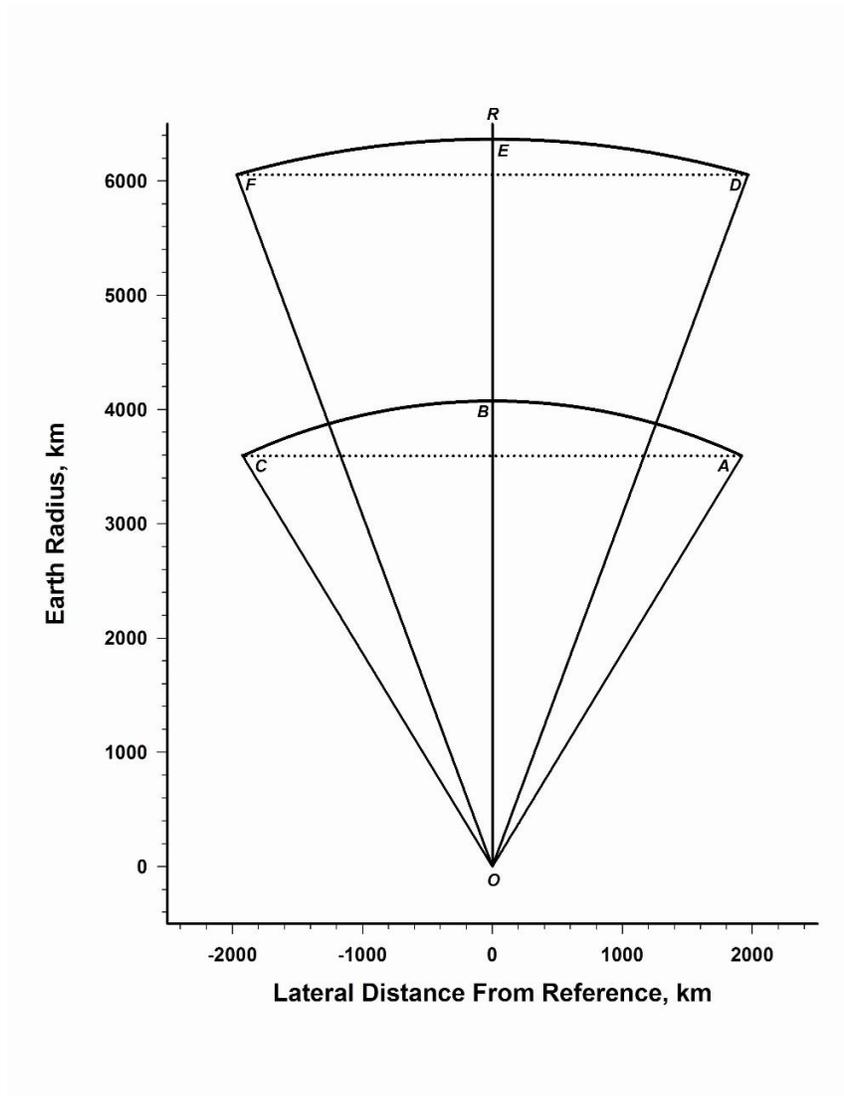

**Figure 1.** Hypothetical 4000 km "ancient" continent cross-section at a radius of 64% of Earth's present radius and the same 4000 km "present" continent cross-section at present-day Earth radius, taken as 6366 km.

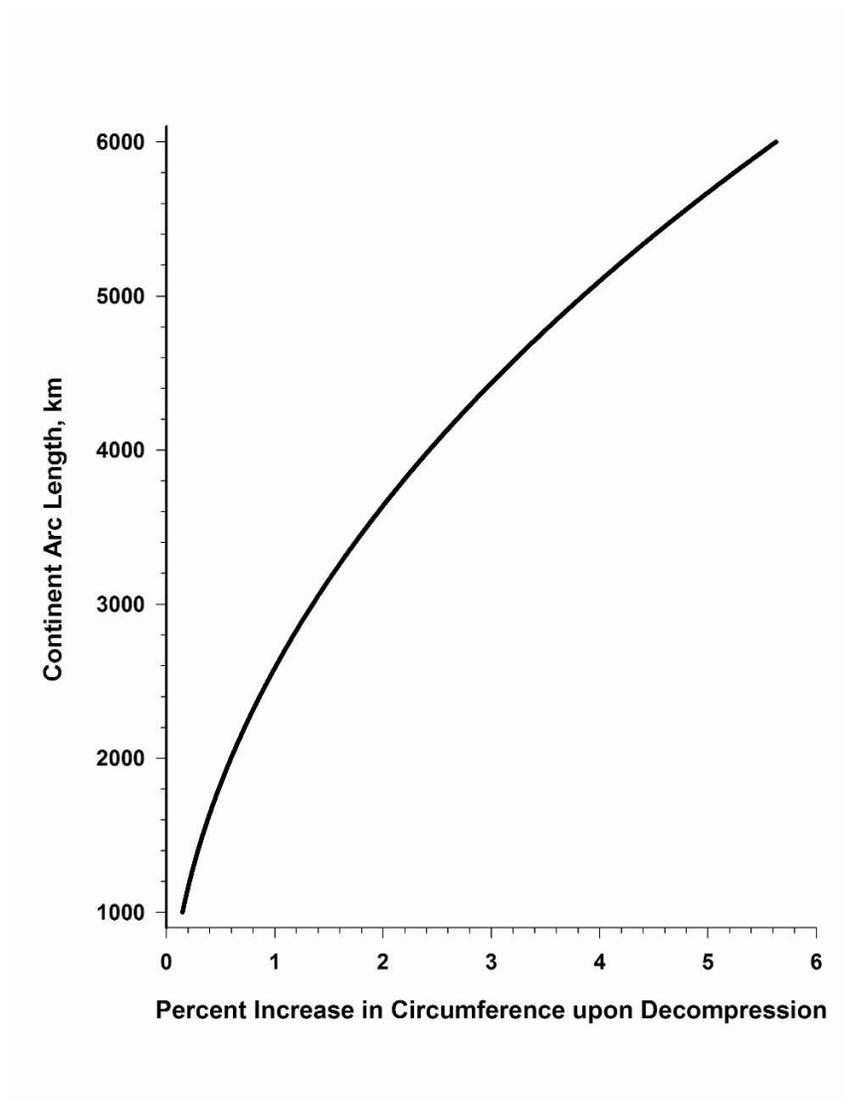

**Figure 2.** Percent increase in peri-continental circumference due to decompression for the hypothetical circular continent shown, if Figure 1, as a function of cross-section arc length.